\documentclass[aip,showpacs,reprint]{revtex4-1}
\usepackage{graphicx,color,epsfig,float}
\usepackage{amssymb,amsmath}

\begin{document}
\title{Asymmetric cluster and chimera dynamics in globally coupled systems}
\author{A. V. Cano}
\affiliation{Grupo de Caos y Sistemas Complejos, Centro de F\'isica Fundamental, Universidad de Los Andes, M\'erida, Venezuela}
\author{M. G. Cosenza}
\affiliation{Grupo de Caos y Sistemas Complejos, Centro de F\'isica Fundamental, Universidad de Los Andes, M\'erida, Venezuela}
\affiliation{School of Physical Sciences \& Nanotechnology, Universidad Yachay Tech, 100119 Urcuqu\'i, Ecuador}

\begin{abstract}
We investigate the emergence of chimera and cluster states possessing asymmetric dynamics in globally coupled systems,
where the trajectories of oscillators belonging to different subpopulations exhibit different dynamical properties.
In an asymmetric chimera state, the trajectory of an element in the synchronized subset is stationary or periodic,
while that of an oscillator in the desynchronized subset is chaotic. 
In an asymmetric cluster state, the periods of the trajectories of elements belonging to different clusters are different. 
We consider a network of globally coupled chaotic maps as a simple model for the occurrence of such asymmetric states 
in spatiotemporal systems. 
We employ the analogy between a single map subject to a constant drive and the effective local dynamics in the globally coupled map system
to elucidate the mechanisms for the emergence of asymmetric chimera and cluster states in the latter system.
By obtaining the dynamical responses of the driven map,
we establish a condition  for the equivalence of the dynamics of the driven map and that of the system of globally coupled maps.
This condition is applied to predict parameter values and subset partitions for the formation of asymmetric cluster 
and chimera states in the globally coupled system. 
\end{abstract}

\date{CHAOS \textbf{28}, 113119 (2018).}

\maketitle

\textbf{Recently a fascinating phenomenon occurring in networks of coupled identical oscillators 
has attracted much attention from researchers in various fields: chimera states. 
A chimera state consists of the simultaneous coexistence of subsets of oscillators with synchronous (coherent) 
and asynchronous (incoherent) dynamics. This behavior represents a state of broken synchronization symmetry and has been  
studied theoretically and experimentally in different contexts and also with a variety of coupling schemes.
In systems with global interactions, chimera states are related to the formation of clusters, where the system segregates 
into distinguishable subsets of synchronized elements.
Here we investigate the emergence of 
chimera states possessing asymmetric dynamics, in the sense that the dynamical evolution 
of oscillators belonging to the synchronized or the desynchronized subset are different: the trajectory shared by the oscillators in
the synchronized subset is stationary, while that of an oscillator in the desynchronized subset is chaotic.
Similarly, we investigate asymmetric cluster states, where the periods of the orbits of oscillators
belonging to different clusters are different. 
In particular, the coexistence of synchronized and desynchronized subsets possessing asymmetric dynamics represents a further breaking
of the synchronization symmetry in a system of coupled identical oscillators.}

\section{Introduction}
There is currently great interest in the investigation of
the emergence of states possessing broken synchronization symmetry in systems of coupled identical oscillators. 
Such behavior, called a chimera state, consists of the coexistence of synchronized and desynchronized subsets of oscillators 
within the system. Initially recognized in networks of nonlocally coupled phase oscillators \cite{Kuramoto,Abrams},
chimera states have also been found in systems with local interactions \cite{Laing1,Clerc,Bera,Hiz} 
and have been investigated in diverse models, including coupled map lattices \cite{Omel,Semenov}, 
Van der Pol oscillators \cite{Ulo}, chaotic flows \cite{Omel2}, 
neural systems \cite{Kanas,Omel3}, quantum systems \cite{Bastidas}, lasers \cite{Rohm}, population dynamics \cite{Dutta}, 
and Boolean networks \cite{Rosin}. Experimental observations of chimera states have been made  
in coupled populations of chemical oscillators \cite{Showalter,Nkomo}, coupled lasers \cite{Hart}, optical light modulators \cite{Roy}, 
electronic \cite{Larger}, mechanical \cite{Martens,Kap,Blaha}, and electrochemical \cite{Kiss} oscillator systems. 
Chimera states may be relevant in real-world phenomena such as the unihemispheric sleep in birds and dolphins \cite{Lima}, 
epileptic seizures \cite{Roth}, neuronal bump states \cite{Sakaguchi}, social systems \cite{JC}, and power grids \cite{Fila}.

Chimera states have been recently found in systems with global interactions \cite{Sen,Pik,Schmidt,Mis,Cano}.
A chimera behavior was observed earlier by Kaneko in a globally coupled map network \cite{Kaneko1}; 
it consisted of the coexistence of one synchronized cluster and a cloud of desynchronized elements. 
These works have revealed that chimera states appear related to the clustering phenomenon typically exhibited by globally
coupled systems, where the system splits into distinguishable clusters of synchronized elements. 

In most reported chimera states, the dynamics of the trajectories of the oscillators in the synchronized or in the desynchronized subsets
are similar; they are both chaotic or both periodic. 
However, regimes where the subsets or clusters exhibit asymmetric behaviors have been seen in systems with long-range or with 
global interactions \cite{Kap,Pik,Schmidt,Ku,Olmi,Kas,Bukh}. 
In this paper, we investigate the emergence of 
chimera states possessing asymmetric dynamics, in the sense that the dynamical evolution 
of oscillators belonging to the synchronized or the desynchronized subset are different: the trajectory shared by the oscillators in
the synchronized subset is periodic or stationary, while that of an oscillator in the desynchronized subset is chaotic.
The coexistence of synchronized and desynchronized subsets with periodic and chaotic dynamics represents a further breaking
of the synchronization symmetry in a homogeneous system. 
Similarly, we study asymmetric cluster states, where the periods of the trajectories of elements from different clusters are different. 

We consider a network of globally coupled chaotic maps as a simple model for the occurrence
of asymmetric chimera and cluster dynamics in spatiotemporal systems.  
We employ the analogy between a single map subject to a constant drive and the effective local dynamics in a globally coupled system of maps
to uncover the mechanisms for the emergence of asymmetric chimera and cluster states in the latter system. 
These asymmetric states can arise in the presence of robust chaos in the local maps.
In Sec.~II we investigate the dynamical responses of a single map driven by a constant
and characterize them on the space of parameters of this system. We establish a condition for the equivalence of the dynamics 
of a steadily driven map and that
of a system of globally coupled maps. In Sec.~III we apply this condition to predict asymmetric chimera and cluster states,
and other collective behaviors, in the globally coupled system. Conclusions are presented in Sec.~IV.

\section{Globally coupled systems}
A global interaction in a system occurs when all its elements are subject to a common influence or field.
A global field may consist of an external source acting on the elements, as in a driven system; or it may originate from
the interactions between the elements, in which case, we have an autonomous system.
As a simple model of an autonomous dynamical system subject to a global interaction, we consider a globally coupled map (GCM) system in the form
\begin{eqnarray}
\label{Global}
{x}^i_{t+1}&=&(1-\epsilon) f({x}^i_t) + \epsilon h_t , \\
h_t&=&\frac{1}{N} \sum_{j=1}^N f({{x}^j_t}).
\label{meanF}
\end{eqnarray}
where $x^i_t$  describes the state variable of the
$i$th map $(i = 1,2,\ldots,N)$ in the system at discrete time $t$; $f$ expresses the local dynamics of the maps;
$h_t$ is a global interaction function, corresponding to the mean field of the system in Eq.~(\ref{meanF}), and the
parameter $\epsilon$ represents the strength of the coupling of the maps to the field. The coupling in Eq.~(\ref{Global})
is assumed in the usual diffusive form.

Several collective states of synchronization can be defined in the system Eq.~(\ref{Global}):

(i) \textit{Synchronization} at time $t$ takes place if $x^i_t=x^j_t$,  $\forall i,j$ in the system. 

(ii) A \textit{desynchronized or incoherent} state occurs when $x^i_t \neq x^j_t$ $\forall i,j$ in the system. 

(iii) \textit{Cluster state}. A dynamical cluster is a subset of elements that are synchronized among themselves. 
In a cluster state, the $N$ elements in the system segregate into $M$ distinct subsets that evolve in time;
i.e., $x^i_t=x^j_t=X^\mu_t$,  $\forall i,j$ in the $\mu$th cluster, with $\mu=1,\ldots,M$.
We call $n_\mu$ the number of elements belonging to the $\mu$th cluster; then its relative size is $p_\mu=n_\mu/N$. 

(iv)  A \textit{chimera state} consists of the coexistence of one or more clusters and a subset of desynchronized elements.
If there are $M$ clusters, the fraction of elements in the system belonging to clusters is $p=\sum_{\mu=1}^M  p_\mu$, while 
the number of desynchronized elements is $(1-p)N$.

A synchronization state at time $t$ can be characterized by
the instantaneous standard deviations of the distribution
of state variables, defined as
\begin{equation}
\sigma (t)= \left[ \frac{1}{N} \sum_{i=1}^N (x_t^i- \bar x_t)^2 \right]^{1/2},
\end{equation}
where
\begin{equation}
\bar{x}_t=\frac{1}{N} \sum_{j=1}^N x^j_t.
\end{equation}

We calculate the fraction of elements that belong to some cluster at time $t$ as \cite{Cano}
\begin{equation}
p(t)= 1-\frac{1}{N}\sum_{i=1}^N \prod_{j=1, j\neq i}^N \Theta \left(|x_t^i-x_t^j|-\delta \right),
\end{equation}
where $\Theta(x)=0$ for $x<0$ and $\Theta(x)=1$ for $x\geq 0$, and $\delta$ is an  appropriate threshold value for 
achieving differentiation between closely evolving clusters. Here we employ $\delta=10^{-6}$.

Statistically, the collective states of synchronization can be characterized through two quantities:
(i) the mean value $\langle p \rangle $ of the fraction of elements that belong to some cluster, 
and (ii) the mean value $\langle \sigma \rangle$ of the standard deviation of the distribution of state variables, 
both obtained by averaging over several realizations of initial conditions and after discarding a number of transients
in each realization \cite{Cano}. Then, a synchronization state corresponds to the values $\langle p \rangle=1$ and  $\langle \sigma \rangle=0$, 
while a cluster state is given by $\langle p \rangle=1$ and $\langle \sigma \rangle>0$. A chimera state is described  by
$p_{\mbox{\scriptsize min}} < \langle p \rangle < 1$ and $\langle \sigma \rangle>0$.  Here we set $p_{\mbox{\scriptsize min}}=0.05$,
as the minimum cluster size to be taken into consideration. An incoherent state possesses values 
$\langle p \rangle < p_{\mbox{\scriptsize min}}$ and $\langle \sigma \rangle>0$.

\begin{figure}[h]
 \includegraphics[scale=.19]{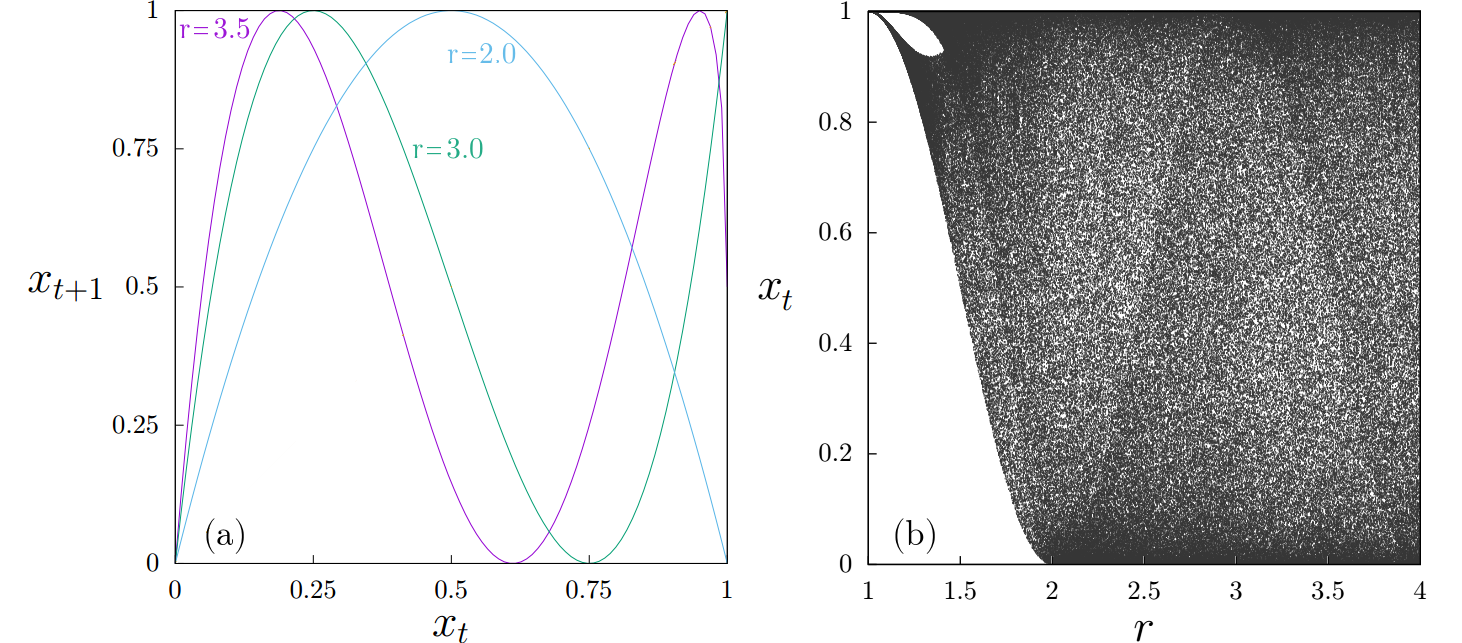}
\caption{(a) Map $x_{t+1}=f(x_t)$ with $f$ in Eq.~(\ref{Aguirre}) for different values of the parameter $r$, as indicated: $r=2$; 
$r=3$; $r=3.5$. (b) Bifurcation diagram of map Eq.~(\ref{Aguirre}) as a function of $r$.}
 \label{f1}
\end{figure}

As local dynamics in the GCM system Eq.~(\ref{Global}), we shall consider the smooth chaotic map \cite{Aguirre}, 
\begin{equation}
\label{Aguirre}
f(x)= \sin^2\left(r\arcsin(\sqrt{x})\right),
\end{equation}
defined on the interval $x \in [0,1]$ for parameter values $r>1$. 
For $r=2$, the map $f$ is unimodal and possesses negative Schwarzian derivative, $Sf <0$. As the parameter $r$ increases, 
the number of maxima of $f$ increases as well, as shown in Fig.~\ref{f1}(a). 
Figure~\ref{f1}(b) shows the bifurcation diagram of the iterates $x_{t+1}=f(x_t)$ of map $f$ as a function of
the parameter $r$. The dynamics exhibits robust chaos with no periodic windows for $r>1$ and fully developed chaos
for $r\geq 2$. The corresponding Lyapunov exponent is $\lambda=\ln r$ \cite{Aguirre}. 

\section{Driven map dynamics}
At the local level, each map in the autonomous GCM system
Eqs.~(\ref{Global}) can be seen as subject to a field that eventually induces a collective state. 
Then, under some conditions, the local dynamics of the GCM system  
should be comparable to that of a single map driven by an external signal in the form \cite{Us1}
\begin{eqnarray}
\label{Driven}
s_{t+1}&=& (1-\epsilon)f(s_t) + \epsilon g(y_t),  \\
y_{t+1}&=&g(y_t).
\label{Gy}
\end{eqnarray}
where $s_t$ is the state of the driven map at discrete time $t$, $f$ is the same function describing 
the local dynamics in Eq.~(\ref{Global}), and the function $g(y_t)$ expresses the influence of the external drive $y_t$. 

In general, an analogy between an autonomous GCM system Eq.~(\ref{Global}) and a driven map Eq.~(\ref{Driven})
arises when the time evolution of the global field $h_t$ is identical to that of 
the drive function $g(y_t)$ \cite{Us1,Orlando}. Then, the corresponding local dynamics in both systems
are similar, and therefore the evolution of any element $x_i^t$ in the GCM system can be equivalent to that of the driven map
for some values of parameters and for appropriate initial conditions. 

Our basic idea is that, by knowing the dynamics of a single driven map, one can infer 
collective behaviors that can appear in a GCM system with similar local dynamics.
This analogy can be applied to
investigate properties induced by the external drive that conduce to specific
cluster or chimera states in an equivalent GCM system. 
A periodic drive function acting on the local map can be associated to the emergence of periodic clusters in a GCM system \cite{Us1}. 
A local map subject to a chaotic drive can be related to chaotic chimera states
(where both synchronized and desynchronized subsets are chaotic) in a GCM system possessing a chaotic global coupling field.
The simplest situation where the driven map analogy can be used arises when the drive function is constant, $g(y_t)= k$; that is,
\begin{equation}
\label{drive}
s_{t+1}= (1-\epsilon)f(s_t) + \epsilon k.
\end{equation}
Then, the equivalence corresponds to a GCM system evolving such that its global field remains constant, $h_t = k$. 
Thus, we shall search for collective states in the GCM system that satisfy this condition.

\begin{figure}[h] 
 \includegraphics[scale=.365]{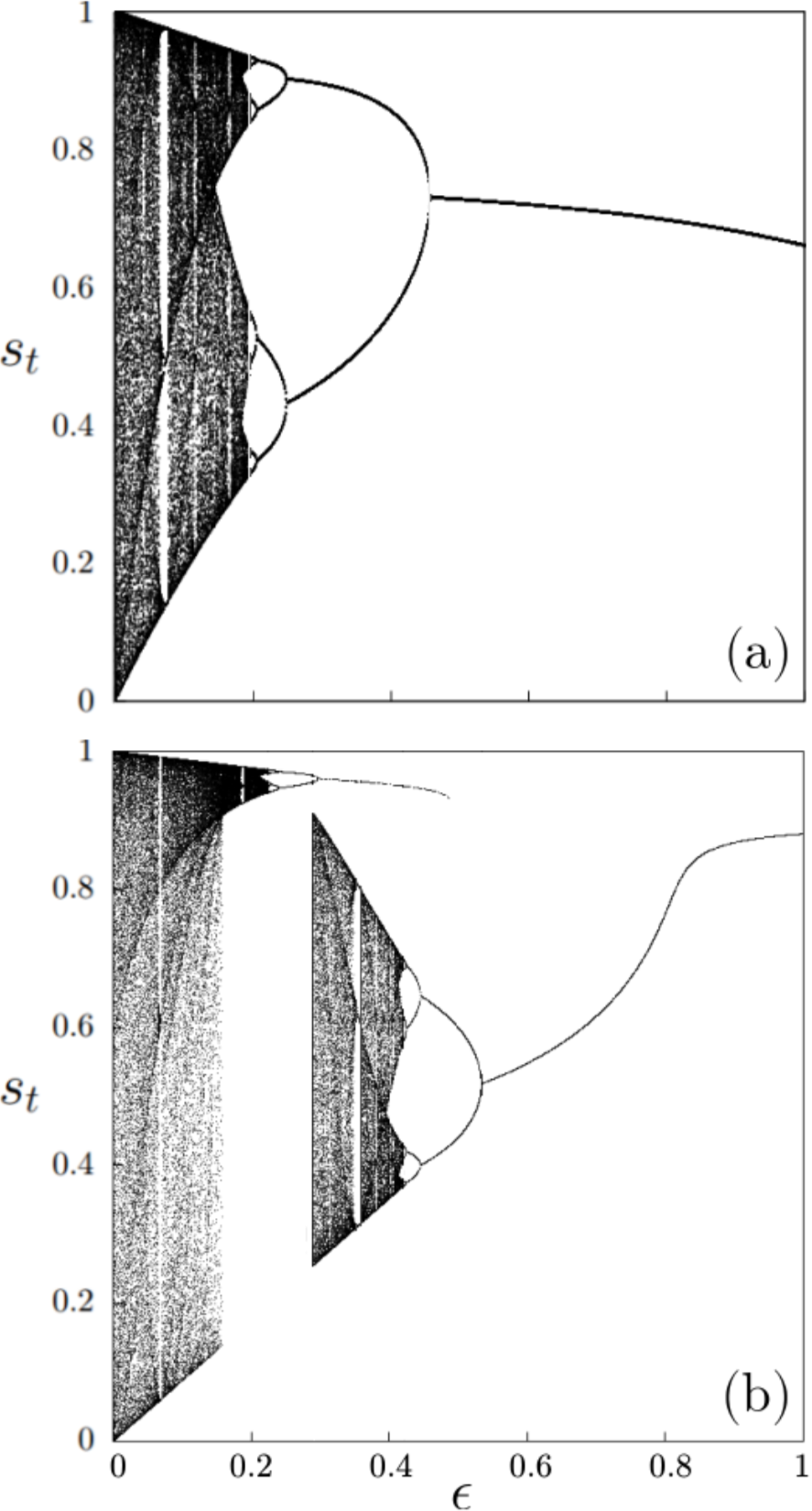}
\caption{Bifurcation diagrams for the driven map Eq.~(\ref{drive}) as a function of $\epsilon$ with fixed values of $r$ and $k$; 
(a) $r=2$, $k=0.66$; (b) $r=3.5$, $k=0.88$.}
 \label{f2}
\end{figure}

Figures~\ref{f2}(a)-(b) show bifurcation diagrams of the driven map $s_t$ in Eq.~(\ref{drive}) as a function of the coupling parameter $\epsilon$
for different values of the parameters $r$ and $k$.
For $r=2$, when the local map $f$ is unimodal, the typical period-doubling bifurcation structure is
observed in Fig.~\ref{f2}(a), which is expected since the driven map $s_t$ is also unimodal and its the Schwarzian derivative is negative.
For $r=3.5$,  $f$ is multimodal and the bifurcation diagram of the driven map Eq.~(\ref{drive} displays period-doubling as well 
as regions of bistability, as seen in Fig.~\ref{f2}(b). The bistability consists of the coexistence of a fixed point with a periodic orbit, 
or the coexistence of a fixed point or periodic orbit with a chaotic attractor. 

In regions where bistability is induced by the drive, different initial conditions $s_0$ can reach different attractors. 
To explore the evolution of different initial conditions, we consider $N$ replicas of the driven map Eq.~(\ref{drive}) or, equivalently, 
a system of $N$ globally driven maps, given by
\begin{equation}
s^i_{t+1} = (1-\epsilon) f(s^i_t) + \epsilon k.
\label{Mdrive}
\end{equation}
where $s^i_t$ $(i = 1,2,\ldots,N)$ describes the state variable of the $i$th replica map in the system at time $t$. 
We can search for states of synchronization emerging in the driven system of Eqs.~(\ref{Mdrive})
analogous to those defined for the autonomous GCM system of Eqs.~(\ref{Global}), 
employing $s_t^i$ variables instead of $x_t^i$. 

\begin{figure}[h] 
\includegraphics[scale=.36]{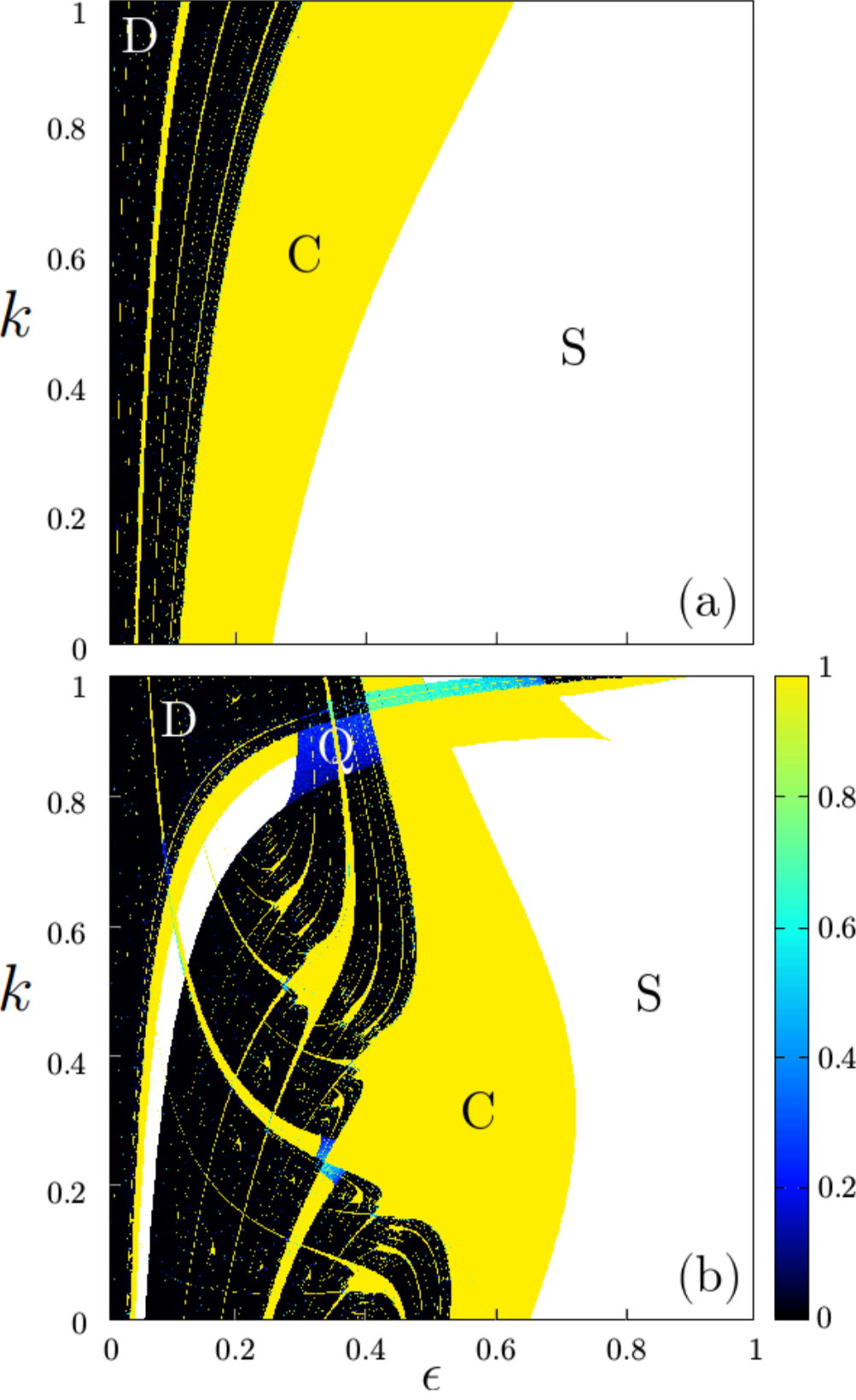}
\caption{Phase diagrams on the plane $(k,\epsilon)$ for the driven system Eqs.~(\ref{Mdrive}) with size $N=1000$ and fixed $r$;
(a) $r = 2$; (b) $r = 3.5$.  
For each data point we calculate the mean values $\langle p \rangle$ and $\langle \sigma \rangle$ 
by averaging over $100$ realizations of initial conditions,
after discarding $10^4$ transients in each realization. 
For each realization, initial conditions $s^i_0$  are randomly and uniformly distributed on the interval $[0,1]$.
The values of $\langle p \rangle$ are indicated by a color code; white corresponds to synchronization.
Labels identify regions of collective states: S, synchronization; C, cluster states; Q, asymmetric chimera states; D, desynchronization.}
 \label{f3}
\end{figure}

Figures~\ref{f3}(a)-(b) show the collective behavior of the driven map system Eqs.~(\ref{Mdrive}) 
on the space of parameters $(k,\epsilon)$, characterized through the quantities $\langle p \rangle $ and $\langle \sigma \rangle$.
Figure~\ref{f3}(a) corresponds to the parameter value $r=2$ for which the map $f$ is unimodal. We observe regions of parameters
where synchronization, cluster states, and desynchronization occur.
In Fig.~\ref{f3}(b), corresponding to the parameter value $r=3.5$ for which $f$ is multimodal,
the system of Eqs.~(\ref{Mdrive}) exhibits synchronization, cluster states, desynchronization,
as well as chimera states for some regions $(k,\epsilon)$.
These behaviors are associated to the appearance of a fixed point, periodic windows, chaos, and bistability, respectively, in 
the bifurcation diagrams of the driven local map, as seen in Figs.~\ref{f2}(a)-(b).

Suppose that a GCM system Eq.~(\ref{Global}) reaches a state such that the evolution of its global field $h_t$ remains constant, $h_t=k$.
In this case, the dynamics of an element $x_t^i$ in the GCM can be described by a single map subject to a constant force $k$, Eq.~(\ref{drive}), and 
the GCM system should be analogous to a system of $N$ driven maps Eqs.~(\ref{Mdrive}); i. e., $x_t^i=s_t^i$. 
The values $s_t^i$ depend on the  parameters $\epsilon$, $k$, and $r$.
Thus, the local dynamics of an autonomous GCM system and the driven map can be equivalent if condition $h_t(s_t^i(r,\epsilon,k))=k$ 
is satisfied. This is the simplest expression of the equivalence between these two systems.
For the mean field $h_t$, this condition is
\begin{equation}
\label{cond}
\frac{1}{N}\sum_{i=1}^N s_t^i(r,\epsilon,k) = k,
\end{equation}
where the possible values of $s_t^i(r,\epsilon,k)$ are the iterates $s_t$ of the single driven map for parameters $(r,\epsilon,k)$. 

\section{Asymmetric  cluster and chimera dynamics in globally coupled maps}
Given a period-$M$ orbit $\{s_1^*,s_2^*,\ldots,s_M^*\}$ in the dynamics of the single driven map,
different initial conditions $s_0^i$ can lead to different out of phase realizations of this orbit.    
Then, a periodic cluster state consisting of $M$ clusters, 
each of period $M$ and moving out of phase with respect to each other, can emerge in the driven system of maps Eqs.~(\ref{Mdrive}).
In general, the formation of periodic cluster states in the driven system of Eqs.~(\ref{Mdrive}) is 
related to the presence of a unique periodic attractor for given parameter values in the single driven map. 
A steadily driven map Eq.~(\ref{drive}) will have a unique asymptotic orbit whenever the local map $f$ is unimodal and $Sf<0$, according to 
Singer's theorem. This is the case for the parameter value $r=2$. 

On the other hand, for values of $r$ such that $f$ is multimodal and the driven map Eq.~(\ref{drive}) exhibits multistability,
there can appear cluster and chimera states in the driven system of Eqs.~(\ref{Mdrive}) possessing partitions with asymmetric dynamical behavior.
In an asymmetric cluster state, maps belonging to different clusters have dynamical trajectories with different periods. 
The occurrence of such a state in the driven system of Eqs.~(\ref{Mdrive}) requires the coexistence of two or more periodic 
attractors with different periods in the dynamics of the single driven map. 
In an asymmetric chimera state, an element from the synchronized subset describes a fixed point or periodic orbit while an element
from the desynchronized subset moves chaotically. This state arises from 
the coexistence of a fixed point or periodic attractor and a chaotic attractor in the driven map. 

Equation~(\ref{cond}) can be employed to predict parameter values or subset partitions for the emergence
of asymmetric cluster or asymmetric chimera states in the GCM system Eqs.~(\ref{Global}) compatible with the condition of a constant mean field. 
We focus on such asymmetric states in this paper.

\subsection{Periodic cluster state}
For a periodic cluster state, the condition $h_t=k$ takes place in the GCM system  when $M$ clusters, each having $N/M$ elements
in a period-$M$ orbit  $X_t^\mu=\{X_1,X_2,\ldots,X_M\}$, are evolving out of phase with respect to each other in order to yield
a constant value $k$ for the mean field $h_t$. For this state, condition Eq.~(\ref{cond}) becomes
\begin{equation}
\label{rc}
 \frac{1}{M} \sum_{j=1}^M s_j^*(r,\epsilon,k) =  k.
\end{equation}
where $\{s_1^*,s_2^*,\ldots,s_M^*\}$ are the points on a unique period-$M$ orbit in the single driven map.
As an example, let $s_1^*(\epsilon,k),s_2^*(\epsilon,k)$ be the points on 
the period-two window in Fig.~\ref{f2}(a) with parameter $r=2$. Then, Eq.~(\ref{rc}) gives
\begin{equation}
\label{2c}
 \frac{1}{2} \left[ s_1^*(\epsilon,k) +s_2^*(\epsilon,k)\right] = k.
\end{equation}
For $k=0.66$ both $s_1^*(\epsilon,k)$ and $s_2^*(\epsilon,k)$ are obtained as functions of $\epsilon$ from the bifurcation diagram
in Fig.~\ref{f2}(a). Then Eq.~(\ref{2c}) can be solved numerically for $\epsilon$. This yields the value $\epsilon=0.29$ for which 
a cluster state comprising two equal-size clusters in period-two, out-of-phase orbits $\{X_1=s_1^*, X_2=s_2^*\}$ such that $h_t=0.66$, 
emerges in the GCM system.

\subsection{Asymmetric cluster state}
Consider an asymmetric cluster state in the GCM system composed of a fraction of $p$
elements in one cluster on a fixed point and a fraction of $1-p$ elements distributed in 
$M$ identical-size, out-of-phase, period-$M$ clusters such that $h_t=k$.  
For this behavior to take place in the GCM system, condition~(\ref{cond}) becomes
\begin{equation}
\label{ac}
p s^* + \frac{(1-p)}{M} \sum_{j=1}^M s_j^*(r,\epsilon,k) =  k,
\end{equation}
where $\{s_1^*,s_2^*,\ldots,s_M^*\}$ are the points in a period-$M$ orbit coexisting with a fixed point $s^*$ 
of the single driven map Eq.~(\ref{drive}).
As an application of Eq.~(\ref{ac}), consider the bifurcation diagram of the single driven map with  
$r=3.5$ and for constant drive $k=0.88$
shown in Fig.~\ref{f2}(b). For the coupling parameter value $\epsilon=0.48$ there is bistability between a fixed point $s^*=0.94$ and a
period-two orbit comprising the points $s_1^*=0.4$ and $s_2^*=0.645$. Then, from Eq.~(\ref{ac}) with $M=2$ we get the fraction
\begin{equation}
\label{rho}
p =\frac{2k-(s_1+s_2)}{2s_*-(s_1+s_2)},
\end{equation}
which for the given values of the variables yields $p=0.86$. Thus, an asymmetric three-cluster state composed of one fixed point cluster
of relative size $p=0.86$ and two out-of-phase period-two clusters, each of relative size $0.07$, and such that $h_t=k=0.88$ can emerge in the
GCM system of Eqs.~(\ref{Global}) for parameter values $r=3.5$ and $\epsilon=0.48$.

\subsection{Asymmetric chimera state}
An asymmetric chimera state consisting of a fraction of $p$ maps synchronized on a fixed point $X^*$
and a fraction of $1-p$ desynchronized chaotic maps, evolving with constant $h_t=k$, may arise in the GCM system
if the following condition is satisfied
\begin{equation}
\label{aq}
p s^* + \frac{1}{N} \sum_{j=1}^{(1-p)N} s_j(r,\epsilon,k) =  k,
\end{equation}
where  $s^*=X^*$ is the fixed point and the $s_j$ are iterates of the single driven map belonging to a chaotic attractor coexisting with $s^*$. 
If $p$ is large enough, the sum term expressing the contribution of the desynchronized chaotic orbits 
reaches a mean value with small fluctuations, and condition (\ref{aq}) can be fulfilled with good approximation. 
As an example, from the bifurcation diagram in Fig.~\ref{f2}(b) for $k=0.88$ we choose the bistable behavior occurring at $\epsilon=0.38$,
where the fixed point $s^*=0.95$ and a chaotic band attractor coexist in the single driven map. 
We can roughly approximate the sum term in Eq.~(\ref{aq}) by the quantity $(1-p)\bar s$, where $\bar s=0.545$ 
is the mean value of the iterates in the chaotic band. Then, from Eq.~(\ref{aq}) with the given parameter values, we obtain
\begin{equation}
\label{pho}
p  \approx \frac{k-\bar s}{s^*-\bar s} = 0.82.
\end{equation}
Consequently, for  $r=3.5$ and  $\epsilon=0.38$ the GCM system of Eqs.~(\ref{Global}) can exhibit an asymmetric chimera state
consisting of the coexistence of a subset staying on the fixed point $X^*=0.95$ and a subset of desynchronized
chaotic maps having approximate relative sizes $0.82$ and $0.18$, respectively, and such that the corresponding mean field reaches 
a constant value.

\subsection{Spatiotemporal patterns for asymmetric cluster and chimera states in globally coupled maps}
\begin{figure}[ht]
\includegraphics[scale=.29]{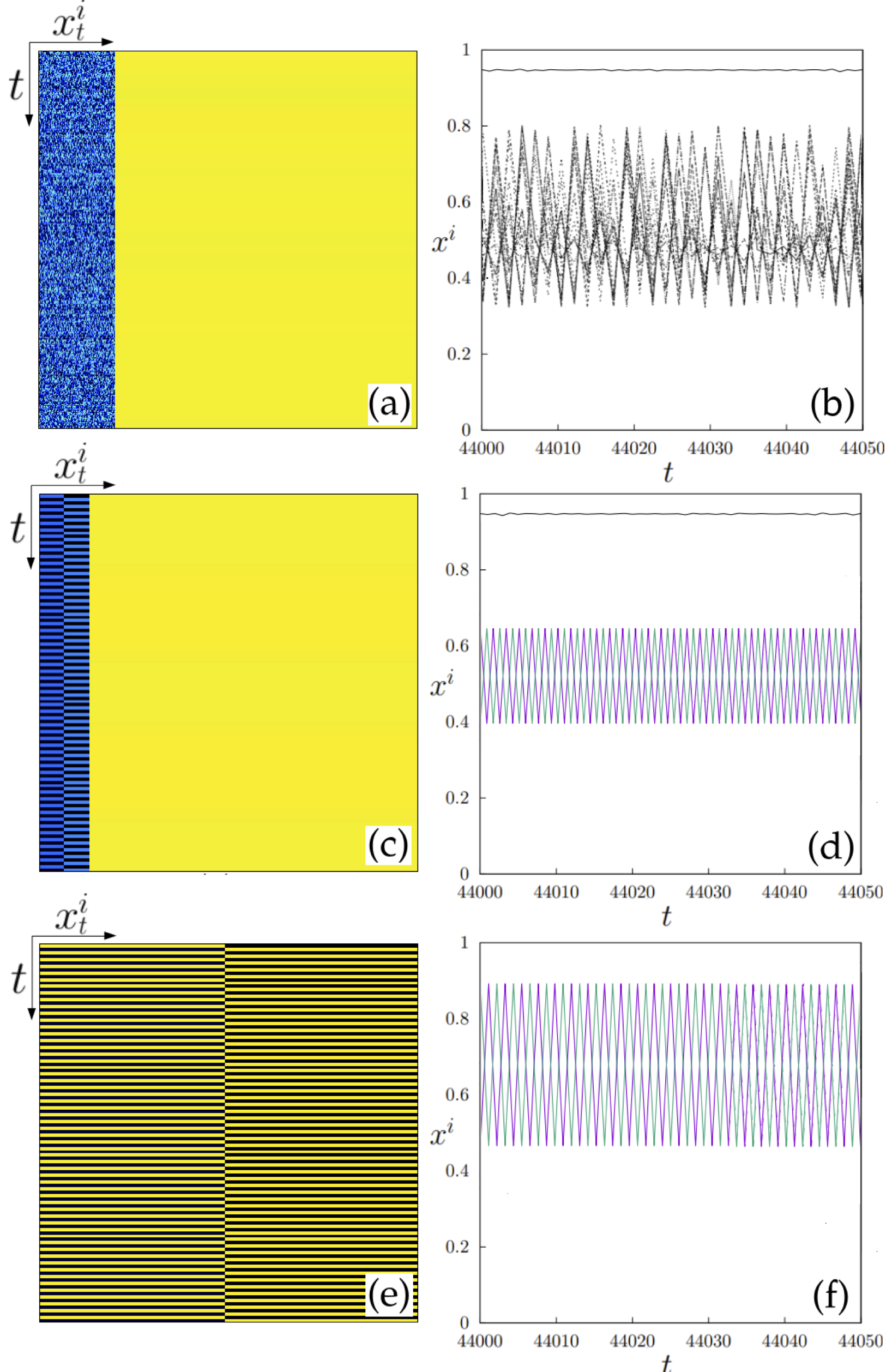}
\caption{Left column: Asymptotic evolution of the states $x^i_t$ (horizontal axis) as a function of time (vertical axis)
for the GCM system of equations (\ref{Global}) with size $N=1000$, for different values of the parameters  $r$ and $\epsilon$. 
For visualization, the indexes $i$ are assigned at time $t=10^4$ such that $i<j$ if $x^i_t < x^j_t$
and kept fixed afterward.
The values of the states $x^i_t$ are represented by the same color coding used in Fig.~\ref{fn5}. 
Initial conditions $x_0^i$ are randomly and uniformly distributed in the interval $[0,1]$.  
After discarding $3 \times 10^4$ transients, $1000$ iterates $t$ are displayed. 
Right column: Time evolution of the states of maps belonging to different subsets corresponding to the patterns on the left.
(a)  $r=3.5$ and $\epsilon=0.38$, asymmetric chimera state; 
(b) state of one map from the synchronized subset (upper curve) and states of 
a few randomly selected maps from the desynchronized subset (lower curves), versus time.
(c)  $r=3.5$ and $\epsilon=0.48$, asymmetric three-cluster state; 
(d) state of one map from the fixed point cluster (upper curve) 
and states of two maps, each from a different period-two cluster (lower curves) versus time.
(e)  $r=2$ and $\epsilon=0.29$, two-cluster, period-two state; 
(f) states of two maps, each from a different period-two cluster, versus time.}
 \label{f4}
\end{figure}

Note that the predictions for periodic clusters, asymmetric clusters and asymmetric chimera states 
possessing constant mean field in the autonomous GCM system
are made solely from the knowledge 
of the dynamical responses of the single driven map and without direct numerical simulation on the GCM system of Eqs.~(\ref{Global}).
Equations (\ref{2c}), (\ref{ac}), and (\ref{aq}) only tell which of those states are possible; they
do not indicate what initial conditions
in the GCM system will conduce to those particular states.
As it is typical of cluster and chimera states in systems of coupled oscillators,  
the predicted states in the autonomous GCM system depend on initial conditions. 
\\ Figure~\ref{f4} shows the spatiotemporal patterns of the variables $x_t^i$ in the autonomous GCM system Eqs.~(\ref{Global}) (left column)
and the corresponding time evolution of states of selected elements from different subsets for each pattern (right column), 
for different values of the local and the coupling parameter.
We employ initial conditions $x_0^i$ randomly and uniformly distributed in the interval $[0,1]$.  
\\ Figure~\ref{f4}(a) shows an asymmetric chimera state for $r=3.5$ and $\epsilon=0.38$, consisting of a subset
synchronized on the fixed point $X^*=0.95$ (with small fluctuations) and a desynchronized subset whose respective
relative sizes are $0.8$ and $0.2$, close to the predicted approximate values. 
The corresponding mean field reaches an almost constant value $h_t = 0.88$ (there are small fluctuations). 
The evolution of the state of one map from the synchronized subset and the states of
four maps from the desynchronized subset are shown in Fig.~\ref{f4}(b).
Figure~\ref{f4}(c) confirms the existence of an asymmetric three-cluster state for parameters $r=3.5$ and $\epsilon=0.48$
with exactly the predicted characteristics, relative sizes, and constant $h_t=0.88$.  
Figure~\ref{f4}(d) shows the corresponding time evolution state of one map from the fixed point cluster
and the states of two maps, each from a different period-two cluster. 
A two-cluster, period-two state appears in Fig.~\ref{f4}(e) for the  parameter values $r=2$ and $\epsilon=0.29$, as predicted. 
The corresponding evolution of the states of two maps, each from a different period-two cluster, is displayed in Fig.~\ref{f4}(f).

\begin{figure}[ht]
\includegraphics[scale=.3]{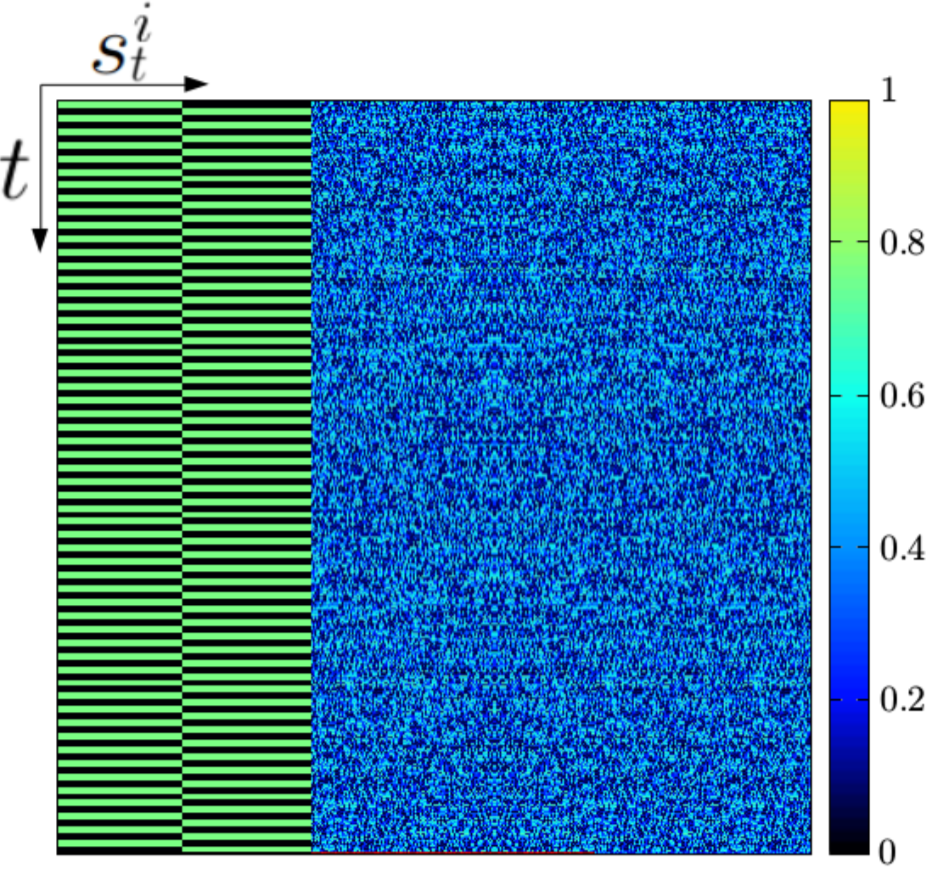}
\caption{Asymptotic evolution of the states $s^i_t$ (horizontal axis) versus time (vertical axis) for the driven
system of maps Eqs.~(\ref{Mdrive}) with size $N=1000$, showing a two-cluster asymmetric chimera state at parameters $r=3.5$,
$\epsilon = 0.34$ and $k=0.26$; a state not occurring in the GCM system.
The color coding representing the values of the states $s^i_t$ is shown. 
Initial conditions $s_0^i$ are randomly and uniformly distributed in the interval $[0,1]$.  
After discarding $3 \times 10^4$ transients, $1000$ iterates $t$ are displayed.}
 \label{fn5}
\end{figure}

Condition Eq.~(\ref{cond}) can also tell what states present in phase diagram $(\epsilon, k)$ of the driven system of maps Eq.~(\ref{Mdrive})
are not possible in the autonomous GCM system. For example, consider the asymmetric chimera state for the driven system shown in
Fig.~\ref{fn5}, consisting of two equal-size, out-of-phase, period-two clusters and a desynchronized chaotic subset. It arises
for parameter values $r=3.5$, $\epsilon=0.34$, and $k=0.26$ in the phase diagram of Fig.~\ref{f3}(b).
Such behavior can be sought in the GCM system through condition Eq.~(\ref{cond}), which for this state takes the form
\begin{equation}
\label{aq2}
\frac{p}{2}(s_1^*+s_2^*) + \frac{1}{N} \sum_{j=1}^{(1-p)N} s_j =  k,
\end{equation}
where $s_1^*=0.09$ and $s_2^*=0.6$ are the points on the period-two orbit of each cluster,
$p/2$ is the relative size of each cluster, and $1-p$ is the relative size of the coexisting desynchronized chaotic subset. 
The sum term in Eq.~(\ref{aq2}) can be roughly approximated as $(1-p)\bar s$, where $\bar s=0.47$ is the value of the middle point
of the width of the chaotic band. Then, from Eq.~(\ref{aq2}) we get $p \approx 1.68$.
Therefore, this asymmetric chimera state cannot occur in the GCM for the given parameter values.

\begin{figure}[H]
\includegraphics[scale=.27]{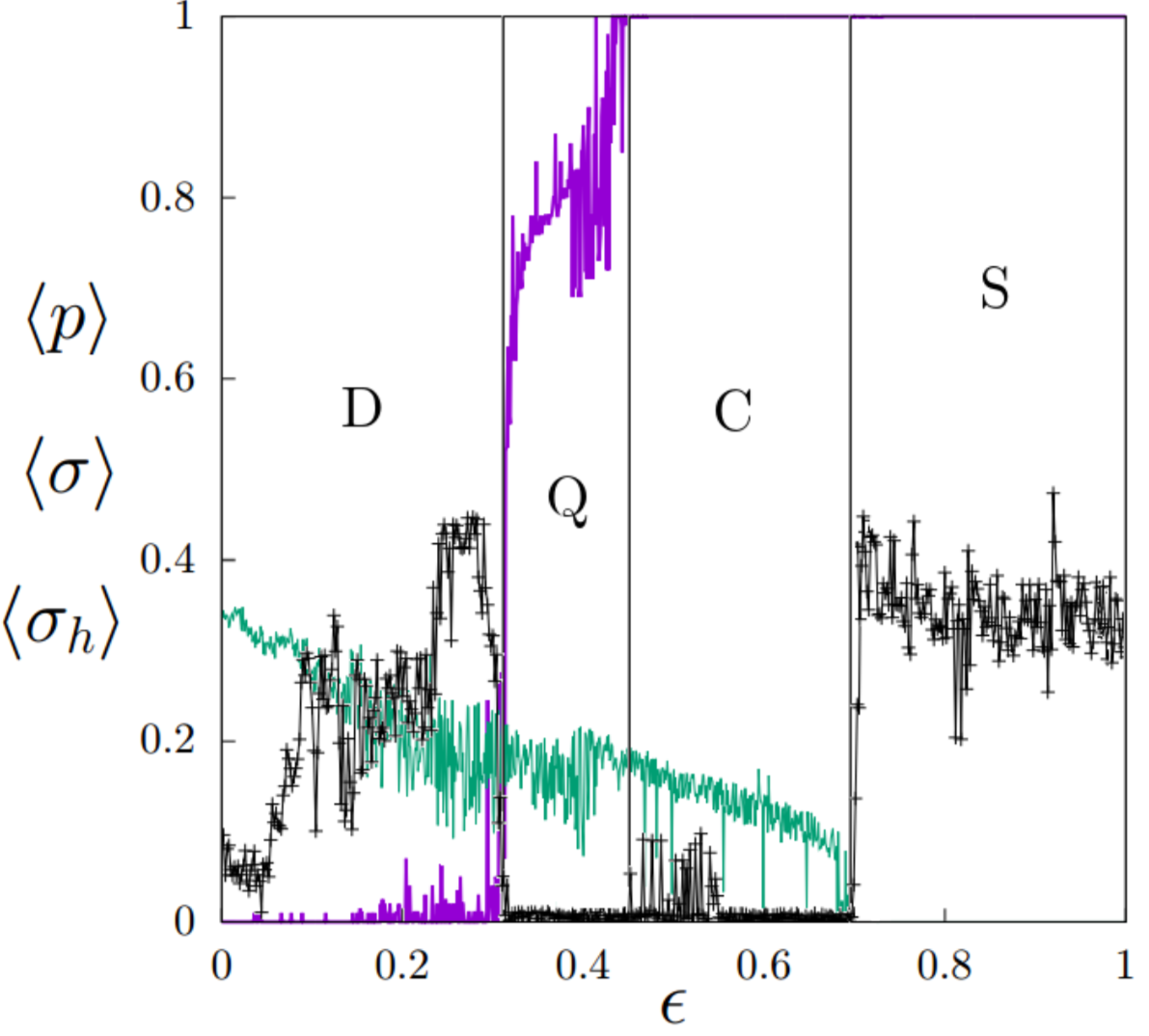}
\caption{Quantities $\langle p \rangle$ (magenta line), $\langle \sigma \rangle$ (green line), and 
$\langle \sigma_h \rangle$ (black line, crosses) as functions of $\epsilon$ for the globally coupled system of 
equations~(\ref{Global})-(\ref{meanF}) with size $N=1000$ and fixed local parameter $r=3.5$. For each value of $\epsilon$ we calculate 
$\langle p \rangle $, $\langle \sigma \rangle$, and $\langle \sigma_h \rangle$ by averaging over $50$ realizations of initial conditions,
after discarding $10^4$ transients in each realization. 
Initial conditions $x^i_0$ are randomly and uniformly distributed on the interval $[0,1]$ for each realization.
Labels indicate collective states: D, desynchronization; Q, asymmetric chimera states; C, cluster states; and S, synchronization.}
\label{f5}
\end{figure}

Asymmetric chimera and cluster states similar to those shown in Fig.~\ref{f4}
emerge in the  autonomous GCM system for a range of values
of the coupling parameter $\epsilon$ corresponding to the region marked $Q$ in Fig.~\ref{f3}(b).
Figure~\ref{f5} shows the quantities $\langle p \rangle$ and $\langle \sigma \rangle$  characterizing the states of 
synchronization as functions of $\epsilon$ for the GCM system of Eqs.~(\ref{Global})-(\ref{meanF}) with fixed $r=3.5$.
Chimera and cluster states appear adjacent to each other for intermediate values of the coupling $\epsilon$. 
To identify those states with asymmetric dynamics, we calculate the standard deviation $\sigma_h$ of
the asymptotic time series of the mean field $h_t$ (after discarding a number of transients). 
A value $\sigma_h=0$ corresponds to a constant field $h_t$ and thus characterizes the asymmetric chimera
or asymmetric cluster states that we are considering.
Figure~\ref{f5} shows the mean value $\langle \sigma_h  \rangle$, obtained by averaging  $\sigma_h$ over $50$ realizations
of initial conditions, as a function of $\epsilon$.  The quantity $\langle \sigma_h  \rangle$ tends to zero over the region of the coupling 
parameter labeled by Q, where chimeras occur, indicating that they correspond to asymmetric chimera states. 
This coincides with the range of $\epsilon$ also labeled $Q$ in Fig.~\ref{f3}(b), 
where asymmetric chimeras arise in the driven system of maps Eqs.~(\ref{Mdrive}). 
In the region labeled by C, both regular clusters and asymmetric cluster states can emerge in the GCM system.
The range of values of $\epsilon$ where synchronization occurs can be calculated from
the stability condition for this state in a system of globally coupled maps in the form of Eqs.~(\ref{Global})-(\ref{meanF}), 
given by \cite{Kaneko1}
\begin{equation}
 |(1-\epsilon)e^\lambda| < 1,
\end{equation}
where $\lambda$ is the Lyapunov exponent for the local map $f$. For the map Eq.~(\ref{Aguirre}) possessing $\lambda=\ln r$, we obtain the
condition $\epsilon > 1-1/r$ for a stable synchronized state. For $r=3.5$, we get the value $\epsilon \geq 0.714$
for synchronization, which agrees with the numerical characterization $\langle \sigma \rangle=0$ for this state labeled S in Fig.~\ref{f5}.

\begin{figure}[ht]
\includegraphics[scale=.29]{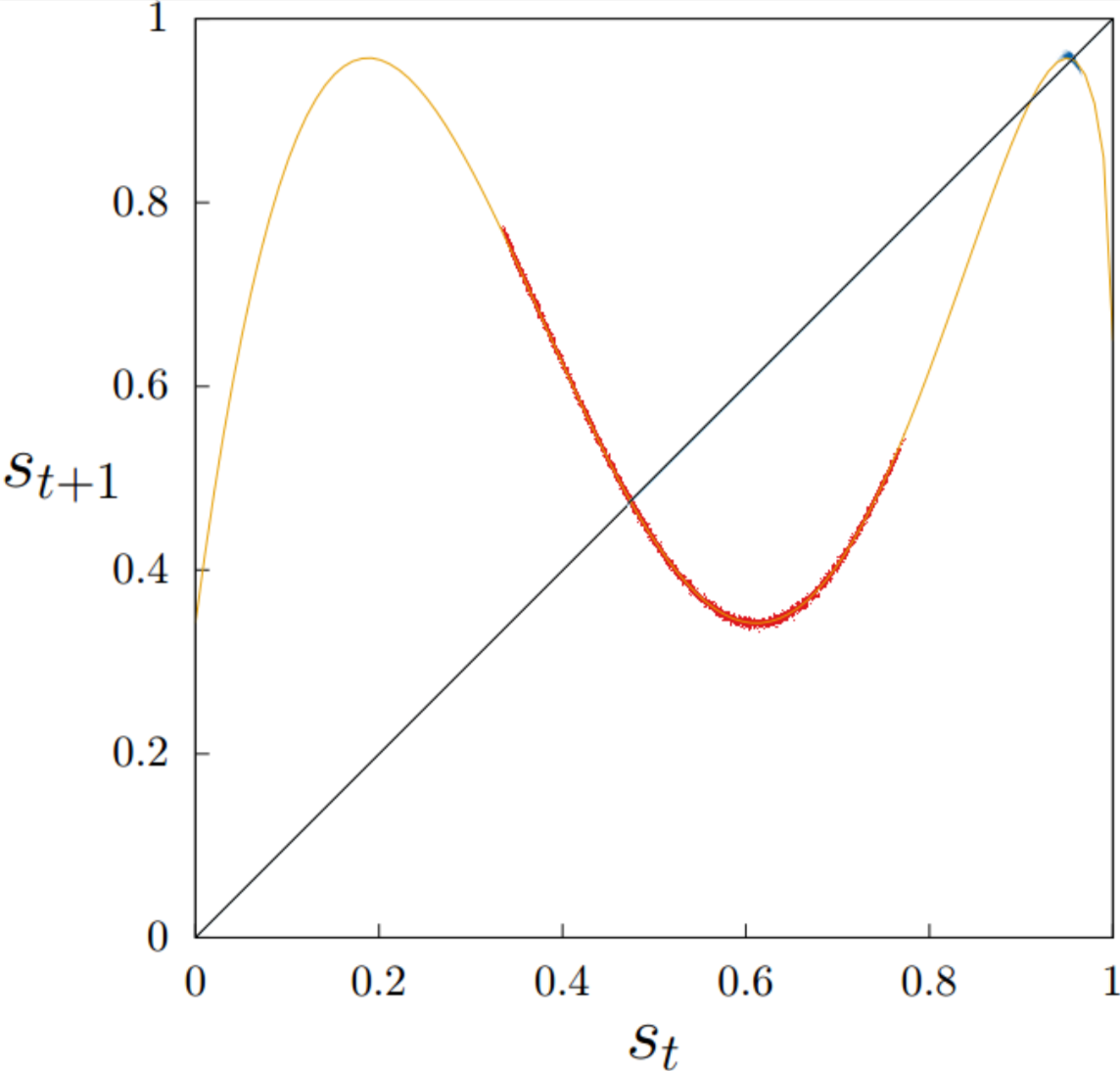}
\caption{Driven map Eq.~(\ref{drive}) for parameters  $r=3.5$, $\epsilon= 0.38$, and $k=0.88$ 
(brown line), and return maps of one element from the synchronized subset (blue dots) 
and one element from the desynchronized subset (red dots) in the asymmetric chimera state of the GCM system 
Eqs.~(\ref{Global})-(\ref{meanF}) shown in Fig.~\ref{f4}(a). The diagonal is also shown.}
\label{f6}
\end{figure}

When $h_t$ becomes constant, the local dynamics of the GCM system can be effectively described by a single map driven by a constant, Eq.~(\ref{drive}). 
The global field produces periodic windows that were absent in the local maps, creating the possibility of multiple out of phase orbits
for the formation of periodic cluster states.
Additionally, the bistability induced by a constant field $h_t$ at the local level explains the emergence of asymmetric cluster and
chimera dynamics in the autonomous GCM system. To visualize this process, 
Fig.~\ref{f6} shows the return maps $x_{t+1}^i$ versus $x_t^i$ corresponding to two elements of the GCM system in the asymmetric chimera state
exhibited in Fig.~\ref{f4}(a) for $r=3.5$ and $\epsilon=0.38$: one element from the synchronized subset 
and another element from the desynchronized subset. The driven map Eq.~(\ref{drive}) with parameters $r=3.5$, $\epsilon=0.38$, and $k=0.88$ 
is also plotted in Fig.~\ref{f6}. Both return maps overlap the driven map. 
The synchronized subset very closely reaches the fixed point attractor $X^*=s^*=0.95$, which is the rightmost fixed point solution 
of the driven map, given by $(1-\epsilon) f(s^*) + \epsilon k=s^*$.
Similarly, the chaotic trajectory of the map from the desynchronized reproduces the dynamics of the chaotic band attractor
coexisting with $s^*$ in the driven map.

\begin{figure}[ht]
\includegraphics[scale=.27]{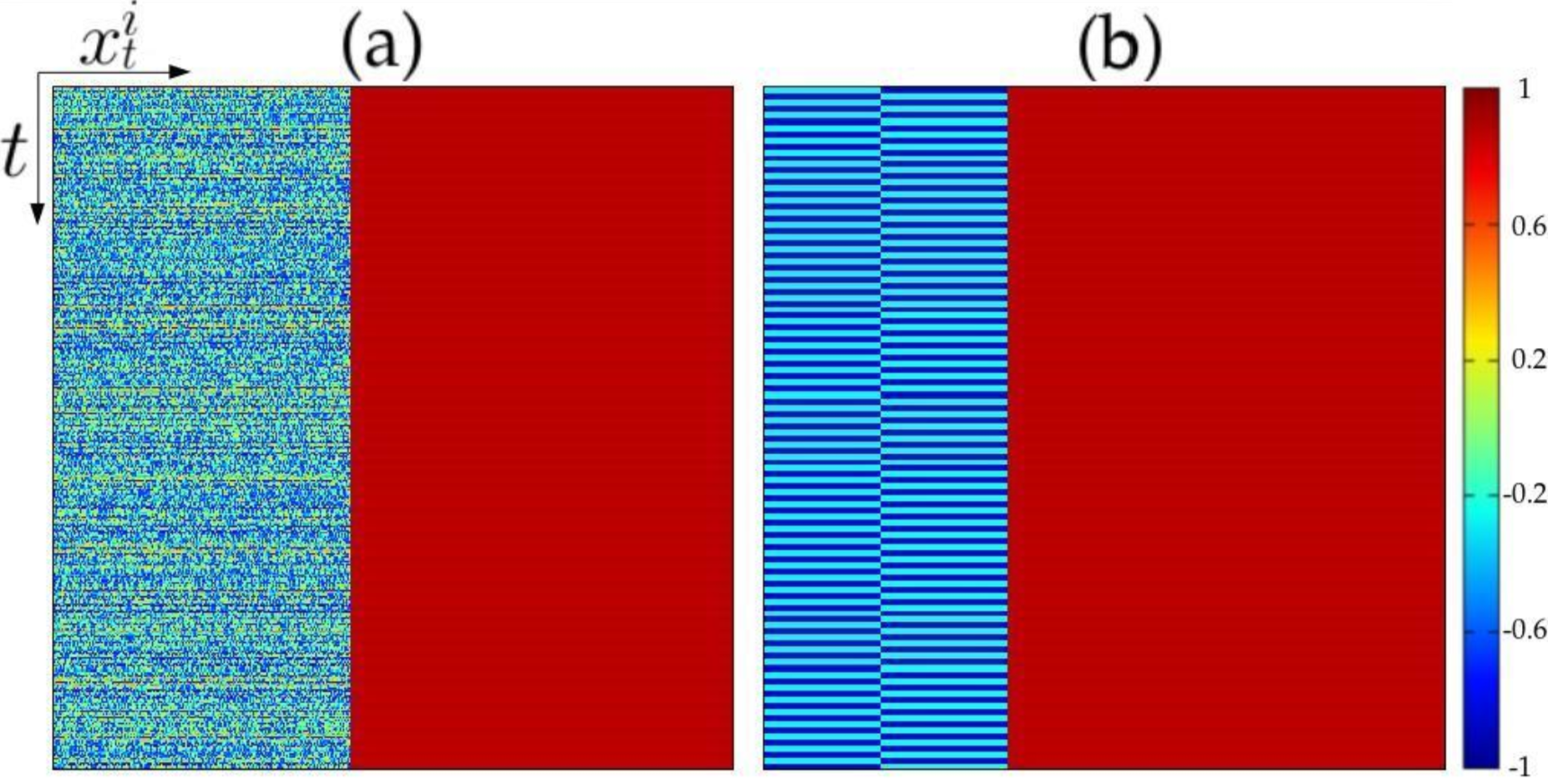}
\caption{Asymptotic evolution of the states $x^i_t$ (horizontal axis) as a function of time (vertical axis)
for the GCM system of equations (\ref{Global}) with  size $N=1000$ and local map $f(x)=1-2(1-2x^2)^2$.
Initial conditions $x_0^i$ are randomly and uniformly distributed in the interval $[-1,1]$.  
After discarding $3 \times 10^4$ transients, $1000$ iterates are displayed. 
The values $x^i_t$ are represented by the color coding right bar.
(a) Asymmetric chimera state, $\epsilon=0.43$. (b) Asymmetric three-cluster state, $\epsilon=0.46$.}
\label{f7}
\end{figure}

Asymmetric cluster and chimera states with patterns similar
to those shown in Fig.\ref{f4} can arise in GCM systems with a different local map $f$,   
as long as $f$ exhibits multistability under constant forcing. For example, the composed logistic map  
$f(x)=q(q(x))$, with $q(x)=1-2x^2$, possesses two maxima and it also displays induced bistability when subject to a constant drive, 
allowing the appearance of asymmetric states with a constant mean field in a globally coupled system of these maps, 
as shown in Fig.~\ref{f7}.  

Note that an equivalence condition can also be established between the local dynamics of an element in a GCM system 
displaying a periodic mean field and a single map subject to a drive having
the same period of the mean field. For a period-$P$ driving function $g(y_t)=\{y_1,\ldots,y_P\}$, the equivalence condition
becomes a system of $P$ equations: $\frac{1}{N}\sum_{i=1}^N s_t^i(r,\epsilon,y_1, \ldots, y_P) = y_j$, $j=1,\ldots,P$. 
In this situation, asymmetric chimera or asymmetric cluster states may arise in the GCM system with unequal partitions
of clusters such that the resulting mean field has period $P$.

\section{Conclusions}
We have investigated the emergence of asymmetric cluster and chimera states in globally coupled systems,
where the trajectories of oscillators belonging to different clusters or subsets exhibit different dynamical properties.

Based on the analogy between the local dynamics of an autonomous GCM system and a single driven map in its simplest form,
when the drive is constant, we have elucidated the mechanisms for the occurrence of cluster and chimera states in systems with global interactions.
The presence of a unique periodic attractor in the dynamics of the single map driven by a constant
can give rise to a periodic cluster state. On the other hand, asymmetric states arise when the dynamics of
the driven map shows bistable behavior; asymmetric cluster states are associated to the coexistence of two attractors
with different periods, while asymmetric chimera states require the coexistence of a fixed point attractor and a chaotic attractor. 
These states can occur in a GCM system evolving such that its global field remains constant. 

By obtaining the dynamical responses of the local map subject to a constant drive, we have established
a condition in Eq.~(\ref{cond}) that links the behaviors of the driven map and a GCM system displaying a constant mean field. 
This condition can be applied to predict parameter values and partition sizes for the occurrence of asymmetric cluster 
and chimera states in the GCM system, or to find out  states allowed by the driven map dynamics that are not possible in the GCM system. 
We have shown that random and uniform distributions of initial conditions, which are commonly used in networks of coupled oscillators,
can lead to the predicted states. In addition, we have 
statistically characterized these states through many realizations 
of random, uniform initial conditions on a range of parameters of the GCM system.

The local map $f$ in Eq.~(\ref{Aguirre}) possesses a homogeneous, single chaotic band attractor. 
Bistability was not present in $f$, nor in the composed
logistic map used in the second example.
The first map is asymmetric and the second is symmetric with respect to the point
$x=0.5$ on the unit interval.
Thus, the occurrence of asymmetric cluster and chimera states cannot be attributed to
preexisting bistability, periodic orbits, or specific symmetry in the local dynamics, nor to
special initial conditions. 
A necessary condition is that the form of the local 
dynamics should be close enough to a bistable behavior
where coexisting attractors possess different dynamical properties,
so that  bistability can be induced  
by a global interaction field, 
either an external driving or an autonomous coupling function.
The emergence of an asymmetric chimera state associated to the bistability produced by the coupling function 
may be seen as a mechanism for self-control of chaos in subpopulations of dynamical elements in an autonomous spatiotemporal system.

Multistability induced by the coupling can also appear in continuous time coupled oscillators \cite{Ujj}.
Our results suggest that asymmetric cluster and chimera states may be induced by an external constant uniform field acting on
an ensemble of chaotic oscillators. By varying the coupling strength or the intensity of the force, 
these asymmetric states could be selected; 
the formation of asymmetric chimeras could be employed as a method for partial or localized control of spatiotemporal chaos. 
Such settings can be experimentally realized and could have applications in diverse systems.


\begin{thebibliography}{99}
\bibitem{Kuramoto} Y. Kuramoto and D. Battogtokh, Nonlinear Phenom. Complex Syst. \textbf{5}, 380 (2002).
\bibitem{Abrams}  D. M. Abrams and S. H. Strogatz, Phys. Rev. Lett. \textbf{93}, 174102 (2004).
\bibitem{Laing1} C. R. Laing, Phys. Rev. E \textbf{92}, 050904(R) (2015).
\bibitem{Clerc} M. G. Clerc, S. Coulibaly, M. A. Ferr\'e, M. A Garc\'ia-\~Nustes, and R. G. Rojas,
Phys. Rev. E \textbf{93}, 052204 (2016).
\bibitem{Bera} B. K. Bera and D. Ghosh, Phys. Rev. E \textbf{93}, 052223 (2016).
\bibitem{Hiz} J. Hizanidis, N. Lazarides, and G. P. Tsironis, Phys. Rev. E \textbf{94}, 032219 (2016).
\bibitem{Omel} I. Omelchenko, Y. Maistrenko, P. H\"ovel, and E. Sch\"oll, Phys. Rev. Lett. \textbf{106}, 234102 (2011).
\bibitem{Semenov} V. Semenov, A. Feoktistov, T. Vadivasova, E. Sch\"oll, and A. Zakharova, Chaos \textbf{25}, 033111 (2015).
\bibitem{Ulo} S. Ulonska, I. Omelchenko, A. Zakharova, and E. Sch\"oll, Chaos \textbf{26}, 094825 (2016).
\bibitem{Omel2} I. Omelchenko, B. Riemenschneider, P. H\"ovel, Y. Maistrenko, and E. Sch\"oll, Phys. Rev. E \textbf{85}, 026212 (2012).
\bibitem{Kanas} J. Hizanidis, V. Kanas, A. Bezerianos, and T. Bountis, Int. J. Bif. \& Chaos \textbf{24}, 1450030 (2014).
\bibitem{Omel3} I. Omelchenko, A. Provata, J. Hizanidis, E. Sch\"oll, and P. H\"ovel, Phys. Rev. E \textbf{91}, 022917 (2015).
\bibitem{Bastidas} V. M. Bastidas, I. Omelchenko, A. Zakharova, E. Sch\"oll, and T. Brandes, Phys. Rev. E \textbf{92}, 062924 (2015).
\bibitem{Rohm} A. R\"ohm, F. B\"ohm, and K. L\"udge, Phys. Rev. E \textbf{94}, 042204 (2016).
\bibitem{Dutta} T. Banerjee, P. S. Dutta, A. Zakharova, and E. Sch\"oll, Phys. Rev. E \textbf{94}, 032206 (2016).
\bibitem{Rosin} D. P. Rosin, D. Rontani, and D. J. Gauthier, Phys. Rev. E \textbf{89}, 042907 (2014).
\bibitem{Showalter} M. R. Tinsley, S. Nkomo, and S. Showalter, Nat. Phys. \textbf{8}, 662 (2012).
\bibitem{Nkomo} S. Nkomo, M. R. Tinsley, and K. Showalter, Phys. Rev. Lett. \textbf{110}, 244102 (2013).
\bibitem{Hart} J. D. Hart, K. Bansal, T. E. Murphy, and R. Roy, Chaos \textbf{26}, 094801 (2016).
\bibitem{Roy} A. Hagerstrom, T. E. Murphy, R. Roy, P. H\"ovel, I. Omelchenko, and E. Sch\"oll, Nature Phys. 8, 658 (2012).
\bibitem{Larger} L. Larger, B. Penkovsky, and Y. Maistrenko, Phys. Rev. Lett. \textbf{111}, 054103 (2013).
\bibitem{Martens} E. A. Martens, S. Thutupallic, A. Fourrierec, and O. Hallatscheka, Proc. Natl. Acad. Sci. USA \textbf{110}, 10563 (2013).
\bibitem{Kap} T. Kapitaniak, P. Kuzma, J. Wojewoda, K. Czolczynski, and Y. Maistrenko, Sci. Rep. \textbf{4}, 6379 (2014).
\bibitem{Blaha} K. Blaha, R. J. Burrus, J. L. Orozco-Mora, E. Ruiz-Beltr\'an, A. B. Siddique, V. D. Hatamipour, 
and F. Sorrentino, Chaos \textbf{26}, 116307 (2016).
\bibitem{Kiss} M. Wickramasinghe and I. Z. Kiss, PloS One \textbf{8}, e80586 (2013).
\bibitem{Lima} N. C. Rattenborg, C. J. Amlaner, S .L. Lima, Neurosci. Biobehav. Rev. \textbf{24}, 817 (2000).
\bibitem{Roth} A. Rothkegel, K. Lehnertz, New J. Phys. \textbf{16}, 055006 (2014).
\bibitem{Sakaguchi}  H. Sakaguchi, Phys. Rev. E \textbf{73}, 031907 (2006).
\bibitem{JC} J. C. Gonz\'alez-Avella, M. G. Cosenza, M. San Miguel, Physica A \textbf{399}, 24 (2014).
\bibitem{Fila} A. E. Filatova, A. E. Hramov, A. A. Koronovskii, S. Boccaletti, Chaos \textbf{18}, 023133 (2008).
\bibitem{Sen} G. C. Sethia and A. Sen, Phys. Rev. Lett. \textbf{112}, 144101 (2014).
\bibitem{Pik} A. Yeldesbay, A. Pikovsky, and M. Rosenblum, Phys. Rev. Lett. \textbf{112}, 144103 (2014).
\bibitem{Schmidt} L. Schmidt and K. Krischer, Phys. Rev. Lett. \textbf{114}, 034101 (2015).
\bibitem{Mis} A. Mishra, S. Saha, C. Hens,  P. K. Roy, M. Bose,  P. Louodop, H. A. Cerdeira, and S. K. Dana, 
Phys. Rev. E \textbf{92}, 062920 (2015).
\bibitem{Cano} A. V. Cano and M. G. Cosenza, Phys. Rev. E \textbf{95}, 030202(R) (2017).
\bibitem{Kaneko1} K. Kaneko, Physica D \textbf{41}, 137 (1990).
\bibitem{Ku} W. L. Ku, M. Girvan, and E. Ott, Chaos \textbf{25}, 123122 (2015).
\bibitem{Olmi} S. Olmi, Chaos \textbf{25}, 123125 (2015).
\bibitem{Kas} D. V. Kasatkin, S. Yanchuk, E. Sch\"oll, and V. I. Nekorkin, Phys. Rev. E \textbf{96}, 062211 (2017). 
\bibitem{Bukh} A. Bukh, E. Rybalova, N. Semenova, G. Strelkova, and V. Anishchenko, Chaos \textbf{27}, 111102 (2017).
\bibitem{Aguirre} J. M. Aguirregabiria,  arXiv:0907.3790 (2009).
\bibitem{Us1} A. Parravano and  M. G. Cosenza, Phys. Rev. E \textbf{58}, 1665 (1998).
\bibitem{Orlando} O. Alvarez-Llamoza and  M. G. Cosenza, Phys. Rev. E \textbf{78}, 046216 (2008).
\bibitem{Ujj} S. R. Ujjwal, N. Punetha, A. Prasad,  and R. Ramaswamy, Phy. Rev. E \textbf{95}, 032203 (2017).
\end{thebibliography}
\end{document}